\documentclass[useAMS]{spie}  
\usepackage[]{graphicx}

\title{Calibration strategies for the LAD instrument on-board LOFT} 

\author{Luigi Pacciani\supit{a,l}, Paolo Soffitta\supit{a}, Andrea Argan\supit{b}, Didier Barret\supit{c}, Enrico Bozzo\supit{d},\\
Marco Feroci\supit{a,l}, George W. Fraser\supit{e}, Jan-Willem den Herder\supit{f}, Martin Pohl\supit{g},\\
Christian Schmid\supit{h}, Chris Tenzer\supit{i}, Andrea Vacchi\supit{j},\\
Dave Walton\supit{k}, Gianluigi Zampa\supit{j},\\
Silvia Zane\supit{k}
\skiplinehalf
\supit{a}IAPS-INAF, Via fosso del Cavaliere, 100 - I-00133, Roma, Italy; \\
\supit{b}INAF Headquarter,  Via del Parco Mellini, 84 I-00136 Roma, Italy; \\
\supit{c}IRAP and Univesit\'{e} de Toulouse,  9, Avenue du Colonel Roche, BP 44346, 31028,  Toulouse Cedex 4, France; \\
\supit{d}ISDC, Chemin d'Ecogia, 16, Versoix, Geneva, Switzerland; \\
\supit{e}Leicester University, University road, Leicester, LE1 7RH,  UK; \\
\supit{f}SRON, Sorbonnelaan 2, NL 3584, CA, Utrecht, The Netherland;\\ 
\supit{g}Istitut f\"{u}r Physik und Astronomie, Universit\"{a}t Potsdam, 14478 Potsdam-golm, Germany; \\
\supit{h}Dr. Karl Remeis-Sternwarte \& ECAP, Universit\"{a}t Erlangen-N\"{u}rnberg, Germany; \\ 
\supit{i}Kepler Center f\"{u}r Astro- und Teilchenphysik  - Institut f\"{u}r Astronomie und Astrophysik, Universit\"{a}t T\"{u}bingen, Sand 1, 72076 T\"{u}bingen, Germany; \\
\supit{j}INFN - Sezione di Trieste, Padriciano 99, I-34127 Trieste, Italy; \\         
\supit{k}Mullard Space Science Laboratory, UCL, Holmbury St Mary, Dorking, Surrey, RH56NT,UK;\\
\supit{l}INFN - Sezione di Tor Vergata, Via della Ricerca Scientifica, 1, I-00133  Rome - Italy
}

\authorinfo{Further author information: (Send correspondence to Luigi Pacciani)\\Luigi Pacciani: E-mail: luigi.pacciani@inaf.it, Telephone: +39 06 4993 4009}

\begin{document} 
  \maketitle 

\begin{abstract}
The Scientific objectives of the LOFT mission, e.g., the
study of the Neutron Star equation of state and of the Strong
Gravity, require accurate energy, time and flux calibration
for the 516k channels of the SDD detectors, as well as the
knowledge of the detector dead time and of the detector
response with respect to the incident angle of the photons.
We report here the evaluations made to assess the calibration
issues for the LAD instrument.
The strategies for both ground and on-board calibrations,
including astrophysical observations, show that the goals
are  achievable within the current technologies.
\end{abstract}


\keywords{X-ray, Calibrations, X-ray Calibration Facilities, Detectors, X-ray sources, LOFT, ESA, Cosmic Vision}
\pagebreak

\section{INTRODUCTION}
\label{sec:intro}  
With its extremely large effective area (4 m$^2$ at 2~keV, 8 m$^2$ at 5~keV, 10 m$^2$ at 8~keV, 1 m$^2$ at 30~keV), the good energy resolution of the Silicon Drift Detectors (SDD \cite{vacchi1991,campana2011,zampa2011,evangelista2012}), and relatively fast detector response time $<$ 5 $\mu$s, 
the LAD\cite{zane2012} onboard LOFT \cite{feroci2012} could perform high time resolution observations of compact objects, for the study of strong-field gravity, black holes mass and spins, and the equation of state of ultradense matter \cite{feroci2010}.\\
In particular, the LOFT mission, will allow for the study of ultradense matter and neutron star structure, via accurate measurements of spin pulsations, neutron
star masses and radii, and through the astroseismological study of crustal oscillations following intense flares from
Soft Gamma Repeaters. Understanding the properties of ultradense matter and determining its equation of state
(EOS) is one of the most challenging problems in contemporary physics. At densities exceeding that of atomic
nuclei, exotic states of matter such as Bose condensates or hyperons may appear; and at even higher densities a
phase transition to strange quark matter may take place.
These densities in the zero temperature regime relevant to these transitions could be investigated only in neutron stars.\\
The other main objective of the mission is the study of strong field gravity close to black holes and neutron stars, via measurements of black holes mass and
spins from time variability and spectroscopy, general relativistic precession, epicyclic motion, quasi-periodic
oscillations (QPOs) evolution and Fe line reverberation studies in bright AGNs.\\

To achieve these goals \cite{kennedy2012}, the absolute calibration accuracy of the LAD must be 10--15\% (which corresponds to a  5--7\% of error on the neutron star
radius determined from the study of the cooling phase of a type I burst).\\
Dead time is also relevant to all sources where we want to perform accurate
characterization of aperiodic phenomena, particularly at high frequency
(1/t$_{dead}$). We need to be able to calibrate the dead time process to an
accuracy better than the precision afforded by the count rates.
Moreover, the dead time should be less than  0.5--1\% at 1 Crab, because with a too large dead time it will be hard to reduce the error.\\
The strong-field gravity, black hole mass and spins will be addressed via the  study of the Fe line of AGNs. Thence an accurate energy calibration
of the LAD is required.
\section{Calibration activities}
To accomplish these requirements, heavy ground and in-flight calibration campaigns are foreseen. The main activities are described in the following sections.\\
The FEE ASICs of the LAD will be equipped with an electronic calibration system, consisting of a test capacitor which could be connected to the inputs of the analog chains for the electric calibration
through a gate.
A calibration charge will be injected to the test capacitor during electronic calibrations. The electronic calibration procedure consists of a set of calibration trains, each train of a specific
calibration charge. The calibration curve will be obtained from the comparison of the injected charge with the response of each analog chain of the FEE ASIC (see [\cite{pacciani2007}] for details in a similar case).\\
The threshold scan is a more sophisticated procedure, comprising an electronic calibration run for each step of threshold (see [\cite{pacciani2008}] for details in a similar case).
The electronic calibration procedure is a fundamental step of the calibration campaigns.
\subsection{Ground Energy calibration, effective area calibration}
In the following, we 	synthetically  summarize the foreseen calibration activities that we plan to perform on-ground for the LAD.
\begin{itemize}
\item pedestal run, electronic calibrations, threshold scan with electronic calibrators of FEE PCB prior to the detector tile integration (at room, but controlled temperature), a dedicated setup is foreseen.
Purpose: functionality, and comparison after the integration with the detector)

\item pedestal run, electronic calibrations, threshold scan with electronic calibrators of FEE PCB after the detector tile integration (at room, but controlled temperature), a dedicated setup is foreseen.
Purpose: functionality, unbonded strips

\item pedestal run, electronic calibrations, threshold scan with electronic calibrators (and with radioactive sources), radioactive sources run of FEE PCB after  the integration of the detector plane of a module (at room, but controlled temperature), a dedicated setup is foreseen.
Purpose: functionality (bonding verification), and comparison with similar calibrations before the integration with the detector,
we will obtain the ratio C$_{feedback}$/C$_{detector}$, the equalization of test capacitances and of test pulse devices (through the comparison with measurements with radioactive sources).

\item pedestal run, electronic calibrations, threshold scan with electronic calibrators of FEE PCB after the integration with the detector plane of a module (at least 2 temperatures in the range of operations of LAD on orbit), a dedicated setup is foreseen.

\item calibration  with radioactive sources at detector level, prior the collimator mounting  (at least 2 temperatures in the range of operations of LAD on orbit)

\item threshold scan with radioactive sources.
Purpose: high statistics energy calibration, and study on the temperature dependency of the system.

\item pedestal run, electronic calibration, threshold scan, radioactive sources acquisition (and source scan along the tile) after collimators plane mounting on the module at room temperature. 
Purpose: functionality, collimator mounting rough test (effective area study through the  radioactive source scan).

\item pedestal run, electronic calibration, threshold scan, radioactive sources acquisition (and source scan along the tile) after collimators plane mounting on the module; runs will be done at least for 2 temperatures
in the range of LAD on orbit.
Purpose: functionality, collimator mounting rough effective area study (through the  radioactive source scan).

\item test at an X-ray facility (with long guide tube to assure an almost parallel X-ray beam, and high X-ray throughput) at 2/3 temperatures  (+ pedestal run, electronic calibration, threshold scan with electronic calibrator, threshold scan with a monochromatic line from the facility test beam).
Purpose: energy calibration, on- and off-axis effective area.
At the level of panel for MARSHALL-XRCF, and of module for facilities with smaller experimental halls.

\end{itemize}

\subsection{In-flight Energy calibration, Effective area calibration}
Calibration activities will be performed in-flight, both using electronic calibrations, both observing astrophysical sources:

\begin{itemize}

\item pedestal run, electronic calibrations, threshold scan with electronic calibrator (at different temperature conditions).
Purpose: functionality, energy calibration, comparison with ground based calibrations 

\item  Cas A Fe line (+pedestal run, electronic calibrations, threshold scan with electronic calibrator).
Purpose: (energy response, on-, and off-axis effective area)

\item  Pb fluorescence lines during observations of empty fields,  L$\alpha$ (10.5~keV) and L$\beta$ (12.6~keV) lines (+pedestal run, electronic calibrations, threshold scan with electronic calibrator).
Purpose: energy response

\item  Crab scan (+pedestal run, electronic calibrations, threshold scan with electronic calibrator).
Purpose: effective area, and off axis behaviour in energy bins (comparison with the Wide Field Monitor-WFM\cite{donarumma2012} to compensate for fine Crab-variability). Due to the flat field of view of WFM, and the use of the same detector technology
(the SDD), we expect that the use of the WFM to make a relative normalization of the LAD on /off-axis affective area will not introduce systematics, WFM single camera S/NR=250 in one day for Crab).

\end{itemize}

A possible strategy  to make effective area absolute calibration is a two fold strategy:
1) On ground, detector tile by detector tile with calibration sources, and at X-ray facilities.
2) Relative detector tiles inter-calibrations/alignment studies will be addressed with the Crab raster scan in-flight.

\subsection{Ground/In-flight Dead Time study}
We are currently studying the strategy and the hardware implementation of the data acquisition system, and the relevant telemetry involved, with the aim to reduce and
simplify the problems related to the dead time correction during scientific analysis. dead time study could be addressed both on-ground, both in-flight. In the following we
give a short list of activities related to this task: 

\begin{itemize}

\item dead time study moving away from detector plane a source, before integration with collimator,  (it is only the study of a piece, FEE+MBEE).
Purpose: dead time study of FEE+MBEE 

\item dead time study of the whole system through an intensive electronic calibration, on ground (at room but controlled temperature), and in-flight.
Purpose: dead time study of the whole system.

\item dead time study in flight, from the study of flux of bright astrophysical sources (e.g., Sco X-1, Cyg X-1).
A dedicated observing mode is foreseen:
a single module with only a minor number of strips enabled to give the source flux with minor dead time effect,
the remaining LAD modules with nominal configuration, with the plain dead time effect. 

\end{itemize}

\pagebreak

\section{Ground calibrations} 

\subsection{Facilities} 

We are currently studying the feasibility of ground calibrations at the Marshall-XRCF X-ray facility, or at European facilities with long guide-tubes, to assure an almost parallel X-ray beam, and high throughput.\\

The X-ray Calibration Facility (XRCF)\footnote{http://optics.nasa.gov/facilities/xraycal.html}
at Marshall Space Flight Center is a very attractive facility for the LAD calibrations, with its guide-tube which is 518 m long, and with a diameter of 1.5 m at exit.
The experimental chamber has a diameter of 7 m and a length of 23 m, and it is thermally controlled in the temperature range -40 $^\circ$C -- +70$^\circ$ C with a maximal spatial variation of  $\pm$1.1$^\circ$C.\\
There are two X-ray sources: 
1)The Electron Impact Point Source (EIPS), which is an Henke-type source which produces X-rays by focusing an electron beam onto a pure or composite metal target. The energy range of EIPS is 0.09-10~keV;
the EIPS flux intensity on the detector ranges from 1 to 500 photons/sec/cm$^2$, depending on the target.
2)The Rotating Anode Source (RAS) which is an 18 kW Rigaku electron impact X-ray generator operating in the 6-40 kV range, for a current ranging from 10 to 450 mA.
There are three different anode materials and three different cathode configurations.
A Double Crystal Monochromator (DCM) is attached to the RAS. The DCM provides a line in the 1-10~keV range.\\

The MPE X-ray test facility PANTER\footnote{http://www2011.mpe.mpg.de/heg/www/Projects/PANTER/main.html} \cite{freyberg2006} at Max-Planck-Institut f\"{u}r extraterrestrische Physik, has a 
guide tube  123.6 m long, 1 m diameter. The experimental chamber has a diameter of 3.5 m, and a length of 12 m.
An open X-ray source with two filter wheels, and a target wheel with 16 different targets, provides 1--10$\times$10$^3$~ph/cm$^2$/s.
A commercial X-ray source provides a couple of lines between 4.5 and 22~keV, and bremsstrahlung continuum up to 50~keV.
A double crystal monochromator provides a line in the range 1.5–-25~keV (including second order).

The X-ray facility at the Ferrara University department \cite{loffredo2004}
works at 6--60~keV and 15--150~keV, beam size at 1~m from source is selected to be 0.3--1.0~mm$^2$,  with a count rate in the experimental hall of 10$^3$ cts/s.
A Double crystal monochromator (double crystal diffractometer), tunable from 7 to 120~keV, provides the X-ray calibration line.
An upgrade is foreseen with a tunnel of $\sim$100 m.\\

The Palermo X-ray Calibration and Testing (XACT) facility \cite{barbera2006} has a
35 m long vacuum line, with a diameter at the beam exit of 1.5 m, to allow for the illumination of an 80~cm sample.
The X-ray source is an Electron impact multi-anode micro focus Manson Model 5, with 6 anodes, and 4 filters. It produces up to 10$^5$ photons/cm$^2$/s at 20 m distance, operating in the 0.1-20~keV range.
A fixed exit double crystal monochromator, tunable in the range 1-30~keV (producing a beam of size 60$\times$30~cm$^2$ at the test chamber) provides the calibration X-ray line. \\

Due to the beam divergence, the effective area calibration will be performed single module by single module, or detector tile by detector tile, depending on the chosen facility. A scan on the beam angle is
foreseen to calibrate the on-, and off-axis effective area. A movement system is needed to tilt
the detector system with respect to the beam axis. During a run, each detector tile will see the beam inclined with a different offset angle with respect to the other detector tiles.
To obtain a coherent set of data, in the scanning for the off-axis effective area (but onlòy in the central part of the field of view), we have to choose a scanning step according to the offset angle of detector tiles
(1$\div$2 arcmin). IN this way we could study the top of the effective area for each detector tile (or for each module, depending on the chosen facility) with respect to the others.
A coarser scan step is foreseen for the outer positions of the LAD field of view. The alignment of the module within LAD
will be reconstructed via the optical study of the alignment of each module under test with respect to the nominal beam axis. 
The comparison of the optical alignment of the whole LAD, with the optical alignment in the X-ray test facility will allow us to determine the overall effective area of the scientific instrument. A further alignment study
will be performed in-flight with the Crab raster scan.

\subsection{Simulations}
In the following we perform some calculation to evaluate the feasibility of ground calibrations at an X-ray facility. We consider here the case of the Marshall-XRCF.\\
The double crystal monochromator system gives a non-uniform beam\cite{swartz1998}, therefore their usage for the LAD effective area study could be difficult. In the following we investigate the usage of the C-continuum mode for
the study of both energy-calibration and of effective area calibration of the LAD.\\

In our evaluations, we refer here mainly to the paper reporting the Chandra HRMA ground-calibration activities \cite{zhao1998}.
They used the EIPS X-ray source with a Carbon anode which produces a continuum spectrum with contamination from the Si-K$\alpha$, W-M$\alpha$, W-M$\beta$  multiplet (1.78~keV),
Ca-K$\alpha$ (3.69~keV), Ti-L$\alpha$ (4.51~keV), V-K$\alpha$ (4.95~keV), Fe-K$\alpha$ (6.40~keV), W-L$\alpha$ (8.38~keV). Zhao and collaborators \cite{zhao1998} made use of these X-ray lines for the energy calibration of
the solid state detectors (SSD,  Canberra 30~mm ultra low energy Germanium detector \cite{camberra_bege}).\\

To perform the evaluation of the feasibility of calibrations at this X-ray facility, we need a rough
estimation of the Fe-line photon flux, and of the continuum flux at the experimental hall:
from that paper (their Fig.~3 and Fig.~4, referring to RUN-ID~116414, for the flat field evaluation), 
we can roughly evaluate the photon flux from the Fe line at the experimental hall with the following formula:
	\begin{equation}
	\label{eq:flux}
F_{Fe\ line} = \frac{peak\_rate_{mon}}{dE_{bin}}\cdot\sqrt{2\pi}\cdot \frac{FWHM_{SSD}}{2.355}\cdot\frac{1}{\epsilon_{ssd}}\cdot\frac{1}{S_{aper}}\cdot\left(\frac{d_{mon}}{d_{exp\ hall}}\right)^2\, ,
	\end{equation}
where:\\
 $peak\_rate_{mon}$ is the Fe line peak rate (continuum subtracted) measured at the monitor system located at a distance of 38 m from the X-ray source (shown in fig.~3 of Zhao and collaborators\cite{zhao1998},
and it corresponds to $\sim$0.03~c/s);\\
$dE_{bin}$ is the bin size of the spectrum shown (1 channel corresponds to $\sim$0.005~keV);\\
$FWHM_{SSD}$ is the SSD typical energy resolution ($\sim$ 0.16~keV at 6.4~keV)\cite{xrcf_phase1};\\
$\epsilon_{ssd}$ is the typical efficiency of SSD at 6.4~keV ($\sim 1)$\cite{mcdermott1997};\\
$S_{aper}$ is the Area of the aperture used for the SSD at the monitor position ($\sim$0.03~cm$^2$);\\
$d_{mon}$ is the distance of the monitor system from the X-ray source ($\sim$38 m);\\
$d_{exp\ hall}$ is the distance of the experiment from the X-ray source ($\sim$538 m).\\
With this method, we evaluated that the Fe-K$\alpha$ line flux at the experimental-hall is $\sim$0.17 photons/cm$^2$/s.\\
Similarly, we evaluated that the X-ray flux of the continuum component of the spectrum at the experimental hall is $\sim$2.3 photons/cm$^2$/s/keV at 6.4~keV
(the measured rate of the continuum component of the spectrum measured with the SSD at the monitor station is $\sim$ 0.07 c/s at 6.4~keV).\\

Alternatively, we can roughly evaluate the Fe-K$\alpha$ line flux at the experimental hall from RUN ID 110542, that was used for the effective area calibration of shell 6 of HRMA.
In this case we have the following formula for the Fe-K$\alpha$ line flux at the experimental hall:
	\begin{equation}
	\label{eq:flux_bis}
F_{Fe\ line} = peak\_rate\_density_{HRMA\ focus}\cdot\sqrt{2\pi}\cdot\frac{FWHM_{SSD}}{2.355}\cdot\frac{1}{\epsilon_{ssd}}\cdot\frac{1}{A_{eff\ shell6}}\, ,
	\end{equation}
where:\\
 $peak\_rate\_density_{HRMA\ focus}$ is the Fe line peak rate (continuum subtracted) divided for the energy bin of the histogram, measured with the SSD placed at the focus of the Shell 6
(shown in fig.~15 of Zhao and collaborators\cite{zhao1998}, and it corresponds at $\sim$80~c/s/keV);\\
$A_{eff\ shell6}$ is the effective area of the shell 6, which is reported in the same figure of the paper of Zhao and collaborators\cite{zhao1998}($\sim$70~cm$^2$ at 6.4~keV).\\
With this other method, we evaluated that the Fe-K$\alpha$ line flux at the experimental-hall is $\sim$0.19 photons/cm$^2$/s, which is in quite good agreement with the other method, considering the rough estimations
of Fe line peak fluxes, and of the other quantities. Similarly, the flux of the continuum component of the spectrum at the experimental hall could be evaluated to be $\sim$1.9 photons/cm$^2$/s/keV (the measured
rate of the continuum component measured with the SSD in the focus of the shell 6 of HRMA is $\sim$130 c/s/keV at 6.4~keV). \\

\begin{figure}
   \begin{center}
   \begin{tabular}{c}
   \includegraphics[width=9cm]{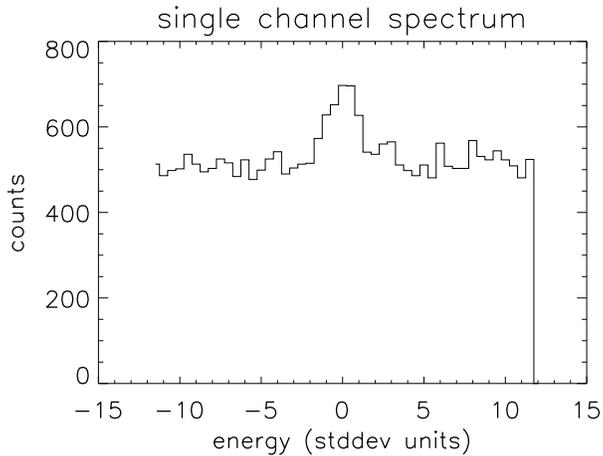}
   \end{tabular}
   \end{center}
   \caption 
   { \label{fig:singlestrip_marshall_spectrum} 
Simulated spectrum for a single SDD channel. Energy units are reported in standard-deviations unit.}
   \end{figure}

\begin{figure}
   \begin{center}
   \begin{tabular}{c}
   \includegraphics[width=12cm]{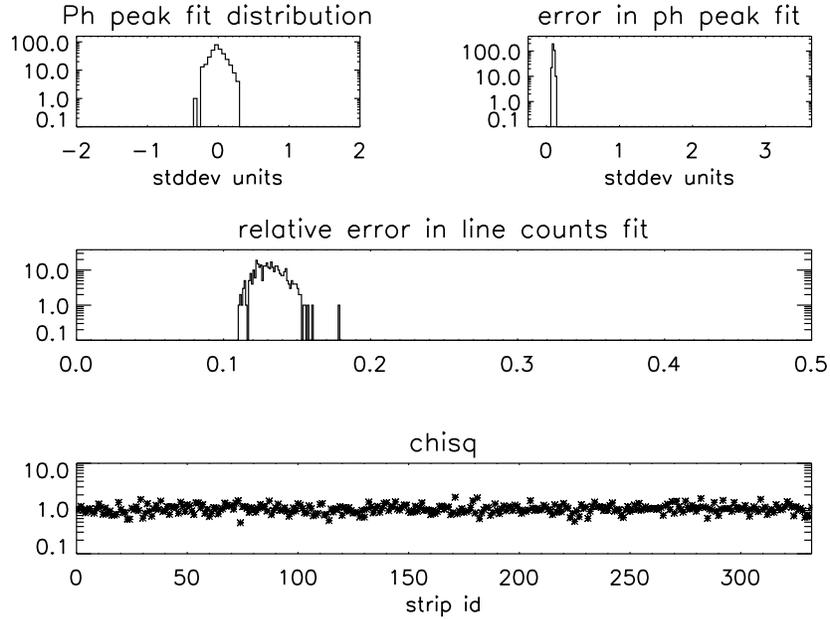}
   \end{tabular}
   \end{center}
   \caption 
   { \label{fig:singlestrip_marshall_fit} 
single strip results of the fit for the energy-calibration assuming an exposure of 113~ks. Top-left panel: distribution of the fitted value of the line peak position (units are reported in standard deviations). Top-right panel:
1-sigma errors in the fit of the line peak position (units are reported in standard deviations). Middle panel: relative error in the counts collected from the X-ray line alone (continuum subtracted).
Bottom panel: $\chi ^2$ of the fit for each of the simulated spectra.}
   \end{figure}

For the case of the LAD ground-calibrations, we simulated a parallel X-ray beam, with a C-continuum like spectrum.
We made use of the spectrum that Zhao and collaborators reported for such a beam \cite{zhao1998}, for the particular case of the Fe-K$\alpha$ line, where we rescaled the line peak to continuum counts/keV, for the
normalization factor $FWHM_{SDD}/FWHM_{Ge\ SSD}$, 
where FWHM$_{SDD}$ is the energy resolution (FWHM) of the SDD detector (0.2~keV).
In Fig.~\ref{fig:singlestrip_marshall_spectrum} we show the simulated spectrum for one of the SDD channels.
The results of the fit are reported in Fig.~\ref{fig:singlestrip_marshall_fit}.\\
To obtain a stable fit, with an error of $\sim$0.1 standard deviations (of the energy resolution)  we need at least 980 counts for each strip.
Assuming a fraction of single strips clusters of 0.4, an efficiency of 0.8 at 6.4~keV, an open fraction 0.5, an SDD strip area of 0.3~cm$^2$, we need an exposure $\sim$113~ks to perform the energy-calibration (strip by strip)
for 1/3 of a LAD panel (the spectrum contains also the other X-ray lines that we need for the energy-calibration procedure). At Marshall-XRCF, we need 3 runs to illuminate a panel, and 18 runs to illuminate the whole LAD.
Thence we need 3.5 weeks of data taking to assure an energy calibration with  an error budget of 1/10 of the LAD energy resolution.\\
 
On- and off-axis effective area calibrations will be less demanding in terms of exposure: following the procedure reported in [\cite{zhao1998}],
we will collect 2500 counts (for an energy bin of 0.1~keV around the Fe line, using only single strip clusters) on each detector tile in 880 s, corresponding to a statistical uncertainty of 1\% per module.
Thence we will calibrate the on-axis effective area (and the effective are for single strip cluster / multiple strips clusters event-topologies) of the whole LAD with an exposure of 14000 s.\\

Additional sources of errors in the measurements come from the evaluation of the quantum efficiency of the monitor system, and from the error in the knowledge of the aperture area of the monitor system (0.8\% for the Chandra HRMA calibrations). 

We note, however, that these evaluations are performed assuming the set-up of the Marshall-XRCF used for the HRMA calibrations. We can act on the EIPS source set-up to assure for an higher X-ray flux,
and shorter calibration exposures.
To further reduce the facility usage, the energy-calibrations could be evaluated making use of the data from the runs for the on- and off-axis effective area calibrations (using the runs for the scan of
the central part of the field of view).\\

We stress here that the beam uniformity at the experimental hall must be investigated for the particular case of the LAD calibrations, requiring a very large and uniform X-ray beam.\\

The usage of a shorter X-ray facility, e.g., the PANTER, will give similar exposures to calibrate the effective area at the level of detector tile, because the shorter distance of the experimental hall from the X-ray source
will allow for higher X-ray fluxes. In this case, for each run, each detector tile will see the source with a different incident angle (with a step of $\sim$ 1 arcmin). Thence the effective area at a given angle at the
level of module will be determined comparing different runs, to construct the effective area with the detector-tiles aligned with the X-ray beam.
\pagebreak
\section{In-flight calibrations}
\subsection{Cas-A SNR}
\begin{figure}
   \begin{center}
   \begin{tabular}{c}
   \includegraphics[width=9cm]{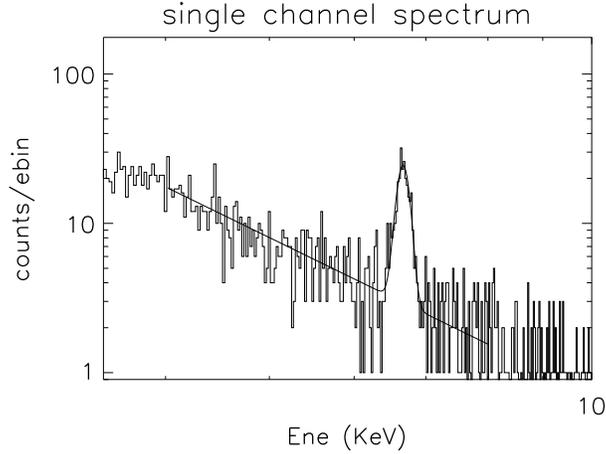}
   \end{tabular}
   \end{center}
   \caption 
   { \label{fig:singlestrip_casa_spectrum} 
     Simulated spectrum for a single SDD channel for a simplified model of Cas A, including a power-law and a Fe Line only.}
   \end{figure}

\begin{figure}
   \begin{center}
   \begin{tabular}{c}
   \includegraphics[width=12cm]{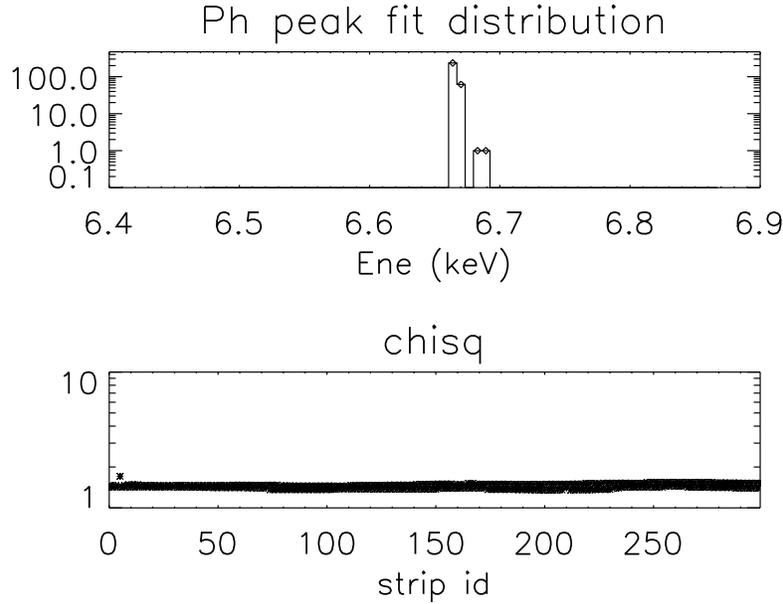}
   \end{tabular}
   \end{center}
   \caption 
   { \label{fig:singlestrip_casa_fit} 
single strip results of the fit for the energy-calibration assuming an exposure of 640ks. Top panel: distribution of the fitted value of the line peak position.
Bottom panel: $\chi ^2$ of the fit for each of the simulated spectra.}
   \end{figure}

A good target for the in-flight calibrations of LAD is the Cas-A SNR, which is a very bright source, with an X-ray luminosity of 80 mCrab (1.5--10~keV) \cite{maccarone1999,maccarone2001}. For our study here, we approximate the spectrum with a power-law of
photon index 3.6, with a superimposed Fe line whose intensity is 7.3$\times 10^{-3}$ ph/cm$^2$/s. The simulated spectrum for a single strip is reported in Fig.~\ref{fig:singlestrip_casa_spectrum}, where
we assume a non earth-occulted fraction of 0.8, resulting in a total observing time of 640~ks.
The distribution of the peak fit for each strip is reported in Fig.~\ref{fig:singlestrip_casa_fit}.
For a good fit we need at least 230 cts/strip from the Fe line, that will be collected in 640Ks of observation (including occultation time).

\subsection{Pb-L$\alpha$ and Pb-L$\beta$ lines from background spectrum}
\begin{figure}
   \begin{center}
   \begin{tabular}{c}
   \includegraphics[width=9cm]{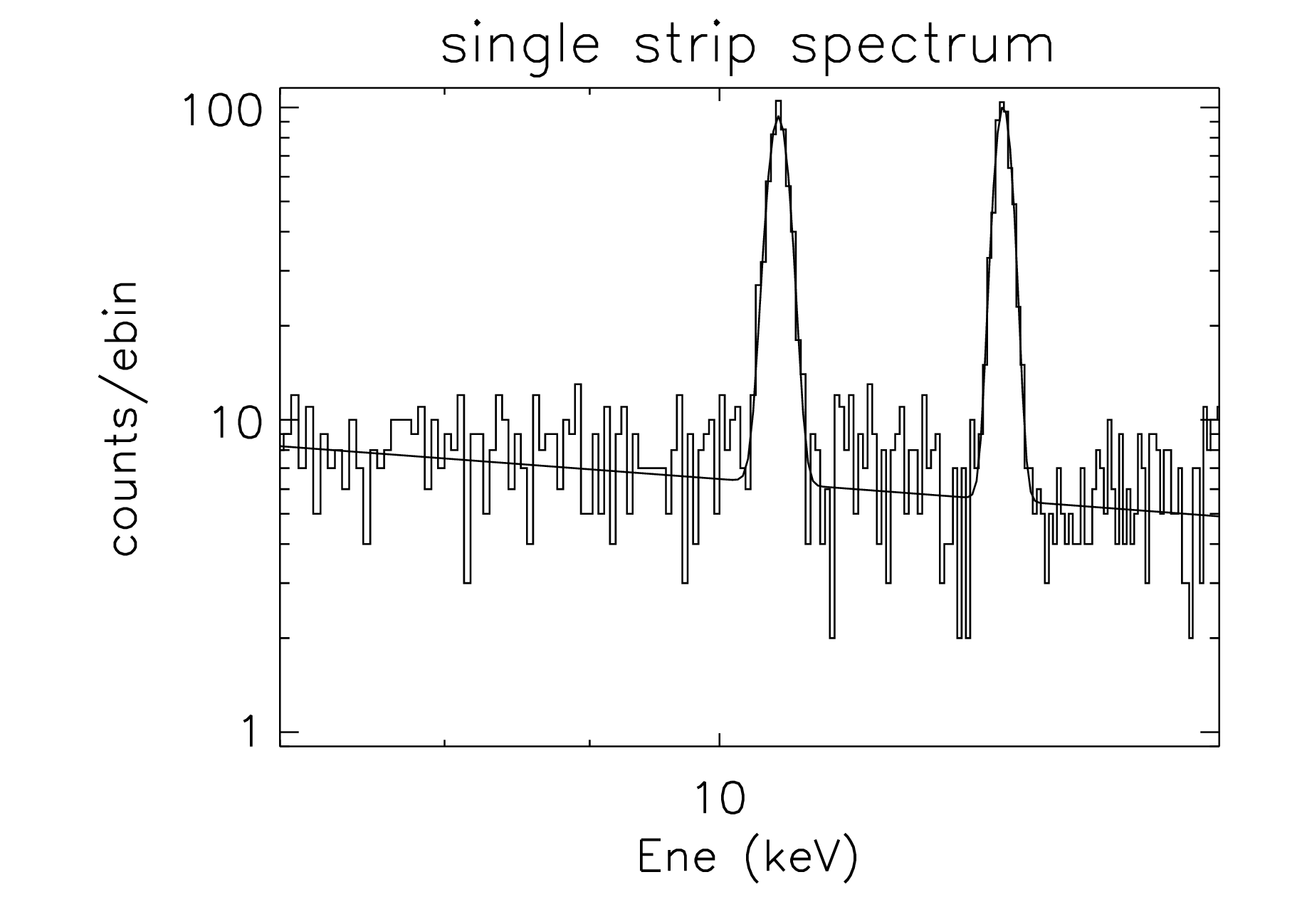}
   \end{tabular}
   \end{center}
   \caption 
   { \label{fig:singlestrip_pbl_spectrum} 
     Simulated spectrum for a single SDD channel for the background spectrum, including Pb-L$\alpha$ and Pb-L$\beta$ fluorescence lines.}
   \end{figure}

\begin{figure}
   \begin{center}
   \begin{tabular}{c}
   \includegraphics[width=12cm]{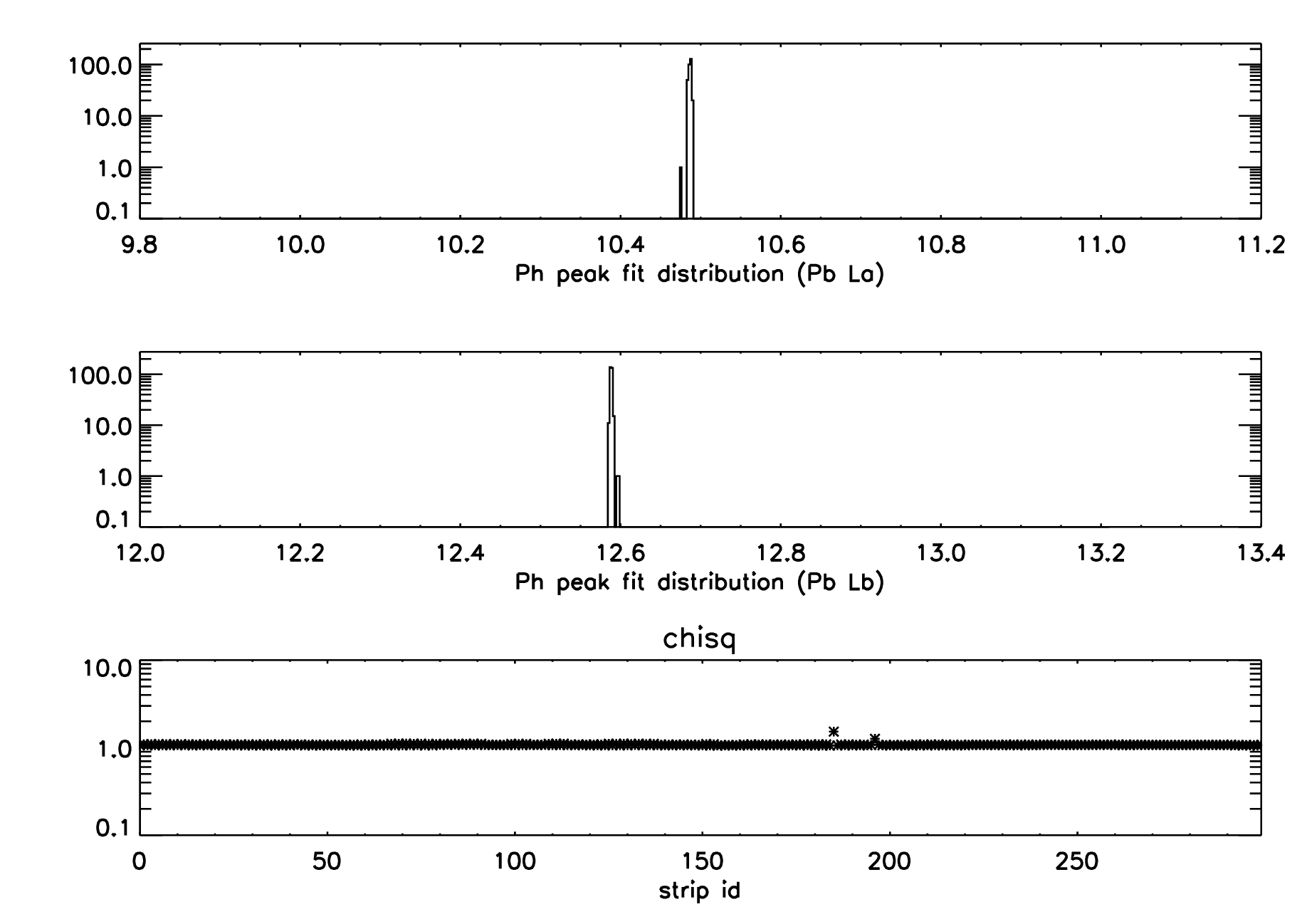}
   \end{tabular}
   \end{center}
   \caption 
   { \label{fig:singlestrip_pbl_fit} 
     Simulated spectrum for a single SDD channel for the background spectrum, including Pb-L$\alpha$ and Pb-L$\beta$ fluorescence lines.}
   \end{figure}

Pb-L$\alpha$ and Pb-L$\beta$ lines will be a bright source for LAD in-flight energy-calibration\cite{campana2012}: we expect a joint Pb-L$\alpha$ and Pb-L$\beta$ lines intensity of
4.6$\times$10$^{-3}$ counts/cm$^2$/s. We simulated the background spectrum corresponding to 680~ks of exposure (Fig.~\ref{fig:singlestrip_pbl_spectrum}).
In the simulation all the events are considered single-anode events.
The distribution of the peak fit for each strip is reported in Fig.~\ref{fig:singlestrip_pbl_fit}.
\section{Conclusions}
We have performed preliminary evaluations of the feasibility of ground and in-flight calibrations. 
This preliminary study shows that the calibrations of the LAD instrument on-board LOFT are an affordable task provided that ground calibrations activities will be performed
at facilities such as the Marshall-XRCF or PANTER-MPE, with time scales of 1 month of facility usage (with the X-ray source set-up and intensity used for the Chandra HRMA
calibrations).\\
The uniformity of the X-ray beam has not been investigated here,but this aspect needs a detailed study, because of the very large area to be calibrated. Also the possibility
of rising the X-ray source intensity has not been covered here, and it surely demands a dedicated effort, with the aim to reduce the facility usage (and its impact on the costs and on
the schedule of the LAD), or to obtain higher statistics.\\

Moreover, a dedicated plan of observations of astrophysical sources, such as Cas-A, and a Crab raster scan are foreseen.
Further in-flight energy-calibrations activities will make use of fluorescence Pb-L$\alpha$ and Pb-L$\beta$ lines produced in the collimation system of the LAD.
The in-flight strategy of calibratons will allow for an independent and complete set of data for the LAD energy and on/off-axis area calibrations.

\end{document}